\documentstyle[twocolumn,aps,epsf]{revtex}
\begin{document}
\draft
%
%
%
%
\title{Frequency-dependence of the Nyquist noise contribution to the
  dephasing rate in disordered conductors}
\author{Axel V\"{o}lker$^{1,\ast}$ and Peter Kopietz$^2$}
\address{$^1$Institut f\"{u}r Theoretische Physik, 
Universit\"{a}t G\"{o}ttingen, Bunsenstrasse 9, 37073 G\"{o}ttingen, 
Germany \\
$^2$Institut f\"{u}r Theoretische Physik, J.W. Goethe-Universit\"{a}t Frankfurt am Main, Robert-Mayer-Strasse 8, 60054 Frankfurt, Germany\\
}
\date{June 11, 2001}
\maketitle
\begin{abstract}
We calculate the Nyquist noise contribution
to the dephasing rate
$1/\tau_{\rm nn}(\omega , T  )$ 
of disordered conductors in $d$ dimensions in the regime
where the 
frequency $\omega$ is larger than the
temperature $T$.  For systems with a continuous spectrum
we find at zero temperature
$1/ \tau_{\rm nn} ( \omega , 0)  \propto \nu_d^{-1} ( \omega / {{D}} )^{d/2}$, 
which agrees qualitatively  with the inelastic quasiparticle scattering rate.
Here $\nu_d$ is $d$-dimensional density of states, and ${{D}}$ is the
diffusion coefficient.
Because at zero frequency and finite temperatures $ 1 / \tau_{\rm nn} ( 0, T )
\propto  [ T / ( \nu_d {{D}}^{d/2} )]^{ 2 / ( 4-d)}$ for $d < 2$, 
the frequency-dependence of  $1/ \tau_{\rm nn} ( \omega , 0)$
in reduced dimensions cannot be obtained by simply
replacing $ T \rightarrow \omega$ in the corresponding finite-temperature
expression for $1/ \tau_{\rm nn} ( 0 , T)$.
We also discuss the dephasing rate
in mesoscopic systems with length $L$ and show that
for $ \omega
\raisebox{-0.5ex}{$\; \stackrel{>}{\sim} \;$}  
{{D}} /L^2$ the spectrum is effectively continuous as far as transport
is concerned. 
We propose weak localization measurements of the AC conductivity
in the GHz-range to clarify  
the origin of the experimentally observed zero-temperature 
saturation of the dephasing rate.

\end{abstract}
\narrowtext

\section{Introduction}
The weak localization correction to the 
DC conductivity of a $d$-dimensional disordered metal 
can be  written as \cite{Altshuler85}
 \begin{equation}
 \delta \sigma = - \frac{e^2 {{D}}}{\pi } 
 \int_{ \tau_{\rm el}}^{\tau_{\phi}} dt ( 4 \pi {{D}} t )^{- d/2}
 \; ,
 \label{eq:dephdef}
 \end{equation}
where $e$ is the charge of the electron, ${{D}}$ is the diffusion coefficient, 
$\tau_{\rm el}$ is the elastic momentum relaxation time, and
$\tau_{\phi}$ is the dephasing time. 
We use units where $\hbar = 1$.
For $d \leq 2$ 
the integral in Eq. (\ref{eq:dephdef}) diverges for $\tau_{\phi} \
\rightarrow  \infty$, so that the temperature-dependence
of the weak localization correction to the
conductivity is sensitive to the temperature-dependence
of the dephasing time $\tau_{\phi}$. In the influential work \cite{Altshuler82}
Altshuler, Aronov and Khmelnitskii showed
that at low temperatures the dephasing rate
$1 / \tau_{\phi}$ in dimensions $d \leq 2$
 is essentially determined by
Coulomb interactions between electrons with
small energy transfers (Nyquist noise).
They found that
below two dimensions the dephasing rate due to 
Nyquist noise is at low temperatures given by \cite{Altshuler82,Voelker00} 
 \begin{equation}
 \frac{1}{\tau_{\phi}} = \left[ \frac{C_d T}{ \nu_d {{D}}^{d/2} } 
\right]^{ 2 / ( 4 -d ) } \; \;  , \; \; d < 2
 \label{eq:tauphi1}
 \; \; ,
 \end{equation}
where $\nu_d$ is the density of states in $d$ dimensions,
and the numerical coefficient $C_d$
diverges for $d \rightarrow 2$, see Ref.\cite{Voelker00}. 
In two dimensions 
the leading temperature-dependence of the dephasing rate is
 \begin{equation}
 \frac{1}{\tau_{\phi} } = \frac{ \ln ( 2 \pi \nu_2 {{D}} ) T}{
 2 \pi \nu_2 {{D}} }
 \; \; , \; \;  d = 2
\; \; ,
 \label{eq:tauphi2}
 \end{equation}
while in $d > 2$ we should write
 \begin{equation}
 \frac{1}{\tau_{\phi} } = 
 \frac{1}{ \tau_{ \rm nn} (T )} + \Gamma_0
 \; \; ,
 \label{eq:tautot}
 \end{equation}
where the Nyquist noise contribution is
 \begin{equation}
 \frac{1}{ \tau_{ \rm nn} (T )} =
 \tilde{C}_d \frac{ T^{d/2}}{ \nu_d {{D}}^{d/2} }
 \; \; , \; \; d > 2 \; .
 \label{eq:tauphi3}
 \end{equation}
An explicit expression for the
numerical coefficient $\tilde{C}_d$  can be found in
Ref.\cite{Voelker00}.
Note that in $d > 2$ the total dephasing rate not only involves
the Nyquist noise contribution $1 / \tau_{\rm nn} $, but 
receives also contributions
from other inelastic processes. In this work we shall only consider the
contribution due to Nyquist noise, and treat all other sources of
dephasing phenomenologically via $\Gamma_0$.

In contrast to the above theoretical predictions,
it was recently found experimentally in a number of
different disordered metals that
the dephasing rate saturates at low temperatures \cite{Mohanty97}. 
Several explanations for the underlying mechanism
of the observed saturation have been proposed 
\cite{Altshuler98,Zaikin98,Imry99,Zawadowski99}. 
However, till now there seems to be no general agreement concerning the
true mechanism that is responsible for the observed
saturation of $\tau_{\phi} ( T) $ at low
temperatures. 
It would  be useful to characterize  the 
various proposed sources for dephasing  in terms of experimentally 
controllable external parameters.
A parameter which can easily be varied is the 
frequency $\omega$ of an external driving field. Note that a
{\it time-dependent} electric field ${ E}_{\rm dri}(t)={
  E}_{ \rm AC} \cos(\omega t)$ gives rise to two
different effects on the dephasing rate. First, the field itself 
directly dephases the wave-functions in non-linear order in 
${E}_{ \rm AC}$. This effect gives rise to a non-equilibrium dephasing rate
$1 / \tau_{\rm AC} ( \omega )$, which has been
calculated by
Altshuler, Aronov and Khmelnitskii \cite{Altshuler81}.
In the regime $  \omega \tau_{\rm AC} ( \omega )  
\raisebox{-0.5ex}{$\; \stackrel{<}{\sim} \;$} 1$ they showed that
 \begin{equation}
 \frac{1}{\tau_{\rm AC} ( \omega ) } \sim  D^{1/5} ( \omega e | {{E}}_{ \rm AC} | )^{2/5}
 \label{eq:tauac}
 \; .
 \end{equation}
Apart from  this direct dephasing,
a frequency-dependent driving field 
leads also to a frequency-dependence of
the dephasing rate due to  Nyquist noise,
so that the total dephasing rate
should be written as
 \begin{equation}
 \frac{1}{ \tau_{\phi}^{\rm tot} ( \omega , T ) } =
 \frac{1}{\tau_{\rm AC} ( \omega ) } + \frac{1}{\tau_{\rm nn} ( \omega , T )} + \Gamma_0
 \; .
 \label{eq:tautotal} 
\end{equation}
Note that for 
$  \omega \gg 1/ \tau_{\rm AC}  ( \omega )$
the direct dephasing rate vanishes as   
$1/ \tau_{\rm AC}  ( \omega ) \propto
\omega^{-2}$ \cite{Altshuler81,Altshuler98}, so that
for sufficiently large $\omega$ the first term 
on the right-hand-side of Eq. (\ref{eq:tautotal})
is negligible. At $T=0$ the dominant dephasing mechanism 
is then due to the frequency-dependence of the
Nyquist noise, i.e. the second term on the right-hand side 
of Eq. (\ref{eq:tautotal}). For $\omega \gg T$ it is
usually assumed that the frequency-dependence 
$ {1}/{\tau_{\rm nn} ( \omega , T )}$
can be obtained by
simply replacing $T \rightarrow \omega $ in the expressions for
$1 / \tau_{\rm nn} ( \omega = 0 , T)$
given above.
In this work we shall show that in dimensions $d \leq 2$ such a procedure
is incorrect.

\section{Perturbative calculation}
\label{sec:diagram}
In this section we calculate $1/ \tau_{\rm nn} ( \omega , T )$
diagrammatically to first order
in the screened Coulomb interaction. We shall adopt
the strategy described in detail by Aleiner, Altshuler and
Gershenson \cite{Aleiner98}, who
focused on $1 / \tau_{\rm nn} ( \omega = 0 , T )$.

The dephasing rate $1 / \tau_{\phi} ( \omega , T )$ for finite
frequency $\omega$ can be defined
by writing the weak localization correction to the
frequency-dependent conductivity $\delta \sigma ( \omega )$ in the form
 \begin{equation}
 \label{wlkorrektur0}
 {\delta\sigma}(\omega) = - \frac{\sigma_d}{\pi \nu_d L^d } 
 \sum_{|{\bf k}|<1/ \ell} {\cal C} ({\bf k} , \omega) \; ,
 \end{equation}
where $\ell$ is the elastic mean free path, and
the Cooperon ${\cal C}({\bf k} , \omega)$ is  given by
\begin{equation}
\label{cooperon1}
{\cal C}({\bf k} , \omega) = \frac{1}{ D {\bf k}^2 -i\omega +1/\tau_\phi} \; .
\end{equation}
Here
$\sigma_d$ is the classical Drude conductivity in $d$ dimensions,
which is related to the $d$-dimensional density of states $\nu_d$ via the Einstein relation $\sigma_d =e^2 \nu_d D$.
 For a truly $d$-dimensional system the volume ${\cal V}_d$ is assumed 
to be ${\cal V}_d =L^d$. The above expression is also valid for quasi 
$d$-dimensional samples embedded 
in three-dimensional space, 
with  a typical transverse extension $a\ll L$ 
and volume ${\cal V}_3 =a^{3-d}L^d$. In this case the density 
of states is given by $\nu_d = a^{3-d}\nu_3 = 1/(\Delta L^d)$, 
where $\Delta$ is the average level spacing at the 
Fermi energy. Note that 
in this case $\nu_3 {\cal{V}}_3 = \nu_d L^d$. 
The rate
$1/\tau_\phi$ in Eq. (\ref{cooperon1}) is the total dephasing rate which 
includes all sources of dephasing. 
To calculate the effect of electron-electron interactions it is assumed
that different dephasing mechanisms are independent, so that 
$1/\tau_\phi$ is simply the sum of the corresponding dephasing rates.
For $ d \leq 2$ 
the electronic contribution is governed 
by the classical Nyquist noise $1/\tau_{\rm nn}$. 
All residual dephasing mechanisms (due to 
external magnetic fields and
electron-electron interactions with large energy transfers)
are taken into account via  
the phenomenological parameter $\Gamma_0$. Hence, the total equilibrium 
dephasing rate is $1/\tau_\phi =  1 / \tau_{\rm nn} + \Gamma_0$. 

If we ignore the Nyquist noise, then the Cooperon is 
\begin{equation}
\label{C0rescal}
{\cal C}_0({\bf k} , \omega) = \frac{1}{ D{\bf k}^2 -i\omega + \Gamma_0} \; .
\end{equation}  
Within a diagrammatic approach
the leading dephasing contributions $\delta\sigma^{\rm deph}$ to 
the weak localization correction 
have been identified by Aleiner {\it et al.} \cite{Aleiner98}.
The relevant diagrams are shown in Fig.\ref{Strom_fig}. 
Evaluating these diagrams we obtain
 \begin{eqnarray}
 \label{sigma_deph3}
 &&\delta\sigma^{\rm deph}(\omega) = 
 \frac{e^2 D}{2\pi^2 i \omega {L}^{2d} } \sum_{\bf
  k,q} \int_{-\infty}^\infty d\omega^{\prime} 
 \nonumber
 \\
 & \times &
 \left\{( \omega^{\prime}+\omega ) [n(\omega^{\prime}+\omega)
  -n(\omega^{\prime})] f_{\bf k}^{{\rm RPA}}(\omega^{\prime} - i 0) 
 \nonumber
 \right.
 \\
 &  &
 \left.
+( \omega^{\prime} -\omega )[ n(\omega^{\prime}-\omega) - 
n(\omega^{\prime})] f_{\bf k}^{{\rm RPA}}(\omega^{\prime} + i 0)
\right\} 
 \nonumber
 \\
 & \times &
\left\{ 2{\cal C}_0({\bf q},\omega^{\prime} + \omega) {\cal C}_0({\bf
    q}, - \omega^{\prime} + \omega )  {\cal C}_0({\bf q-k},\omega) 
\right. \nonumber \\ && \left.
- {\cal C}^2_0({\bf q},\omega) 
 \left[ C_0({\bf q-k},\omega^{\prime} + \omega ) 
 + {\cal C}_0({\bf q-k},- \omega^{\prime} + \omega ) \right]
\right\} \; .
 \nonumber
 \\
 & & 
\end{eqnarray}
Here $n(\omega)$ is the Bose function and 
the screened interaction $f^{\rm RPA}_{ {\bf{q}} } ( \omega \pm i 0)$ 
 is related to the bare Coulomb interaction $f_{\bf q}$ via 
$f^{\rm RPA}_{\bf q}(\omega \pm i 0 ) = f_{\bf q}/\epsilon({\bf q}, \omega \pm i0)$.
In the diffusive regime the dielectric function is given by
 \begin{equation}
 \epsilon ( {\bf{q}} , \omega \pm i0 ) = 1
 + f_{\bf{q}} \nu_d \frac{ D {\bf{q}}^2  }{ D {\bf{q}}^2 \mp i \omega }
 \; .
 \label{eq:epsdef}
 \end{equation}  
The Bose functions 
restrict the range of integration to frequencies $\omega^{\prime}\leq
\max(T,\omega)$. In this regime the effect of the Coulomb interaction
can be described by a fluctuating random field (i.e. Nyquist noise),
which is produced by 
the motion of the electrons in the system.  If
$\Gamma_0\gg 1/\tau_{\rm nn}$ (which can be achieved by applying an
external magnetic field), then Eq. (\ref{cooperon1}) can be expanded in powers
of $1/(\Gamma_0\tau_{\rm nn})$. To first order this yields
\begin{equation}
\label{wlkorrektur1}
\delta\sigma(\omega) = - \frac{\sigma_d}{\pi \nu_d L^d}
\sum_{|{\bf q}|<1/ \ell} \left[{\cal C}_0({\bf q},\omega)
  -\frac{1}{\tau_{\rm nn} }{\cal C}^2_0({\bf q},\omega) \right]\; ,
\end{equation}
where ${\cal C}_0$ is defined in Eq. (\ref{C0rescal}). 
To extract $1/\tau_{\rm nn} ( \omega , T )$ 
from Eq. (\ref{sigma_deph3}), this equation is
rewritten as
\begin{eqnarray}
\label{sigma_deph4}
&& \delta \sigma^{\rm deph}(\omega) = 
 \frac{e^2 D}{2\pi^2 i\omega L^{2d} } \sum_{\bf
  q} {\cal C}^2_0({\bf q},\omega) \sum_{\bf k}\int_{-\infty}^\infty d\omega^{\prime}
 \nonumber
 \\
 & \times &
 \left\{( \omega^{\prime} + \omega ) ][n(\omega^{\prime}+\omega )
  -n(\omega^{\prime})]f_{\bf k}^{{\rm RPA}}(\omega^{\prime} - i 0) 
\right. \nonumber \\ 
 && \left.
+(\omega^{\prime} -\omega)[n(\omega^{\prime}-\omega) - n(\omega^{\prime})]
f_{\bf k}^{{\rm RPA}}(\omega^{\prime} + i 0)
\right\} 
 \nonumber
 \\
 & \times &
\left\{ 2{\cal C}_0({\bf q+k},\omega^{\prime} + \omega ) {\cal C}_0({\bf
    q+k},-\omega^{\prime} + \omega)  {\cal C}^{-1}_0({\bf q},\omega) 
\right. \nonumber \\ && \left.
- {\cal C}_0({\bf q-k},\omega^{\prime} + \omega ) 
- {\cal C}_0({\bf q-k}, -\omega^{\prime} + \omega )
\right\} \; .
\end{eqnarray}
Using $\sigma_d/\nu_d = e^2 D$ and comparing Eqs. (\ref{wlkorrektur1}) and
(\ref{sigma_deph4}), we obtain the following expression for
$1/\tau_{\rm nn} ( \omega , T )$ to first order in the screened interaction
 \begin{eqnarray}
 &&\frac{1}{\tau_{\rm nn}(\omega , T)} =
 \lim_{{\bf q\rightarrow 0}} \frac{1}{2\pi i \omega L^d} 
\sum_{\bf k}\int_{-\infty}^\infty d\omega^{\prime}
 \nonumber
 \\ 
 & \times &
 \left\{( \omega^{\prime} + \omega ) [n(\omega^{\prime} +\omega )
  -n(\omega^{\prime})] f_{\bf k}^{{\rm RPA}}( \omega^{\prime} - i0) 
\right. 
 \nonumber \\ && \left.
+( \omega^{\prime} - \omega ) [
 n(\omega^{\prime}-\omega) - n(\omega^{\prime})]
 f_{\bf k}^{{\rm RPA}}( \omega^{\prime} + i 0)
\right\} 
 \nonumber
 \\
 & \times &
\left\{ 2{\cal C}_0({\bf q+k}, \omega^{\prime} + \omega ) {\cal C}_0({\bf
    q+k}, -\omega^{\prime} + \omega )  {\cal C}^{-1}_0({\bf q},\omega) 
\right. \nonumber \\ && \left.
- {\cal C}_0({\bf q-k}, \omega^{\prime} + \omega ) 
- {\cal C}_0({\bf q-k}, \omega^{\prime} + \omega )
\right\} \; .
\end{eqnarray}
Inserting the explicit form of ${\cal C}_0$, we finally obtain
\begin{eqnarray}
\label{tau_rpa}
\frac{1}{\tau_{\rm nn}(\omega, T)} & =  &
-\frac{1}{\pi i\omega L^d } 
\sum_{\bf k}\int_{-\infty}^\infty d\omega^{\prime}
 \nonumber
 \\
 & & \hspace{-19mm} \times
 \left\{ ( \omega^{\prime}+\omega) [ n(\omega^{\prime}+\omega )
  -n(\omega^{\prime})]  f_{\bf k}^{{\rm RPA}}(\omega^{\prime} - i 0) 
\right. \nonumber 
 \\ &    & \left.  \hspace{-17mm} +
 (\omega^{\prime}-\omega)
 [n(\omega^{\prime}-\omega) - n(\omega^{\prime})]
 f_{\bf k}^{{\rm RPA}}(\omega^{\prime} + i 0)
 \right\} 
 \nonumber 
 \\ 
 &  & \hspace{-19mm} \times 
 \frac{D{\bf k}^2}{ \left[ D{\bf k}^2 -i ( \omega^{\prime} + \omega ) +\Gamma_0 \right]
\left[ D{\bf
    k}^2 + i (\omega^{\prime} - \omega ) +\Gamma_0 \right]} \; .
\end{eqnarray}

\subsection{The dephasing rate for $ T \gg \omega $}

 In the limiting cases $T\gg\omega$ or
$T\ll \omega$ we can evaluate Eq. (\ref{tau_rpa}) analytically.
Let us first consider the limit $T\gg\omega$ and reproduce the
known results \cite{Altshuler82}.
In this limit
the function $n(\omega^{\prime} \pm \omega)$ may be
expanded to first order in $\omega$. The derivative of the 
Bose-function is
\begin{equation}
\label{Bose_deri}
\frac{d n}{d \omega^{\prime}} = - \frac{1}{4 T
  \sinh^2\left(\frac{\omega^{\prime}}{2 T}\right)} \; .
\end{equation}
To evaluate the $\omega^{\prime}$-integral, we approximate
\begin{equation}
\label{Bose_deri_approx}
\frac{1}{ T
  \sinh^2\left(\frac{\omega^{\prime}}{2 T}\right)} \approx \frac{4 T}{
  \omega^{\prime 2}} \Theta\left(|\omega^{\prime}|- T \right) \; ,
\end{equation}
where $\Theta(x)$ is the usual step function. This yields
\begin{eqnarray}
\label{tau_rpaT}
\frac{1}{\tau_{\rm nn}(T)} & = & \frac{2 T}{\pi L^d } 
\sum_{\bf k}\int_{-T}^Td\omega^{\prime} 
 \frac{\text{Im}[f_{\bf k}^ {{\rm RPA}}(\omega^{\prime} - i 0 )]}{\omega^{\prime}}
 \nonumber
 \\
 & \times &
\frac{D{\bf k}^2}{[D{\bf k}^2 -i\omega^{\prime} +\Gamma_0][D{\bf
    k}^2 +i\omega^{\prime} +\Gamma_0]} \; .
\end{eqnarray}
In metals, where the screening is strong and the electrons propagate 
diffusively, the essential contribution is due to momenta $|{\bf k}| \sim
\sqrt{\omega^{\prime}/D}$ with $\omega^{\prime} \ll 1/\tau_{\rm el}$. 
Then we may approximate 
\begin{equation}
\label{f_RPAapprox}
f_{\bf k}^ {{\rm RPA} }(\omega \pm i 0 ) \approx  \frac{D{\bf k}^2
  \mp i\omega}{\nu_d D{\bf k}^2}  \; ,
\end{equation} 
so that
\begin{eqnarray}
\label{tau_rpaT2}
\frac{1}{\tau_{\rm nn}(T)} = \frac{2 T}{\pi \nu_d L^d } 
\sum_{\bf k}\int_{-T}^Td\omega^{\prime} 
\frac{1}{| D{\bf k}^2 -i\omega^{\prime} +\Gamma_0 |^2 } \; .
\end{eqnarray}
For macroscopic samples with a continuous spectrum the above expression
can be calculated by replacing $L^{-d} \sum_{\bf k} \rightarrow \int
d{\bf k}/(2\pi)^d$. In the limit $T\gg\Gamma_0$ one obtains \cite{Aleiner98}
\begin{eqnarray}
\label{tau_fl_pert}
\frac{1}{\tau_{\rm nn}(T)} \sim \left\{
\begin{array}{ll}
 \frac{T}{\nu_1 \sqrt{D \Gamma_0}} \; \; &
 \text{for} \; d=1
\\ 
\frac{T}{2\pi\nu_2 D}
\ln\left(\frac{T}{\Gamma_0}\right) \;\; & \text{for} \; d=2
\\
\frac{T^{3/2}}{\sqrt{2}\pi^2 \nu_3 D^{3/2}}
\;\; & \text{for} \; d=3 \; .
\end{array}
\right.
\end{eqnarray}
A priori these results are only valid if $\Gamma_0\gg
1/\tau_{\rm nn}$. 
However, the correct results for the dephasing rate 
due to Nyquist noise \cite{Altshuler82,Voelker00,Aleiner98} given in
Eqs. (\ref{eq:tauphi1}, \ref{eq:tauphi2}, \ref{eq:tauphi3})
can be obtained by substituting 
$\Gamma_0 \rightarrow  1/\tau_{\rm nn} ( T ) $ 
on the right-hand side of Eq. (\ref{tau_fl_pert}) 
and then solving for $1 / \tau_{\rm nn} ( T )$.

\subsection{Dephasing rate for $ T \ll \omega$}

Eq. (\ref{tau_rpa}) can also be used to obtain the frequency-dependence of
$1/\tau_{\rm nn} (  \omega , T )$ in the limit $\omega \gg T$ 
and $\omega\gg \Gamma_0$. In this case
Eq. (\ref{tau_rpa}) reduces to
\begin{eqnarray}
\label{tau_rpa2}
\frac{1}{\tau_{\rm nn}(\omega)}  & = & -\frac{2}{\pi i\omega L^d } \text{Re}
\sum_{\bf k}\int_{-\omega}^0 d\omega^{\prime} 
 \nonumber
 \\
 & \times &
\frac{  D{\bf k}^2   (\omega+\omega^{\prime})  f_{\bf k}^{{\rm RPA}}
(\omega^{\prime} - i 0 ) }{[D{\bf k}^2 -i 
( \omega^{\prime} + \omega )] [D{\bf
    k}^2 +i ( \omega^{\prime} - \omega ) ]} \; .
\end{eqnarray}
Here the real-part was taken, since only this quantity can be physically
interpreted as a dephasing rate. 
Now the condition $1/\tau_{\rm nn} ( T ) \ll \Gamma_0$ for the validity of the
first order expansion is replaced by 
 \begin{equation}
  \frac{ 1}{\tau_{\rm nn} ( \omega )}  \ll \omega
 \; \; , \; \; \text{(validity of perturbation theory)}
 \; \; .
 \label{eq:condpert}
 \end{equation}
To assess the limitations imposed by Eq. (\ref{eq:condpert}), we should
distinguish the bulk regime from the mesoscopic regime.
The physical properties of the system assume their bulk values
as long as  the dephasing length  $L_{\phi}$ is smaller than the size of the system,
which means that the dephasing rate must be 
larger than the Thouless energy $E_c = D / L^2$, 
 \begin{equation}
 E_c \ll
 \frac{ 1}{\tau_{\rm nn} ( \omega )} 
 \; \; , \; \; \text{(bulk regime)}
 \; \; .
 \label{condbulk}
 \end{equation}
Furthermore, for a mesoscopic system 
the spectrum is effectively continuous if the dephasing rate
exceeds the level spacing $\Delta = ( \nu_d L^{d} )^{-1}$, 
 \begin{equation}
 \Delta \ll \frac{ 1}{\tau_{\rm nn} ( \omega )} 
 \; \; , \; \; \text{(continuous spectrum regime)}
 \; \; .
 \label{condspec}
 \end{equation}
For a good metal $ \Delta \ll E_c$, so that
a bulk metal has always a continuous spectrum. 
Note that in the regime
 \begin{equation}
 \mbox{ max} \{ \Delta , \omega \}  
 \ll \frac{1}{\tau_{\rm nn}( \omega )} \ll
  E_c
 \end{equation}
the system is mesoscopic with a continuous spectrum, and 
the condition (\ref{eq:condpert}) for the validity of
the perturbative identification of $1/ \tau_{\rm nn} ( \omega )$
with the right-hand side of Eq. (\ref{tau_rpa2}) is not satisfied.
In Sec.\ref{sec:eikonal} 
we shall calculate the dephasing rate
via an Eikonal expansion which does not rely 
on a perturbative expansion of the type (\ref{wlkorrektur1}).
We shall show that for a
mesoscopic system with a continuous spectrum
qualitative behavior of the dephasing rate due to Nyquist noise 
resembles that of a bulk system,
\begin{equation}
\label{tau_eps_pert}
\frac{1}{\tau_{\rm nn}(\omega)} \propto
\frac{1}{\nu_d}\left(\frac{\omega}{D}\right)^{{d}/{2}} \; ,
\end{equation}
see Eq. (\ref{tau_rpa2}).
Note that this implies that the condition
(\ref{eq:condpert}) for the validity of perturbation theory becomes
$\omega^{1-d/2}\gg \Delta/E_c^{d/2}$. 
Because $E_c\gg \Delta$ for a good metal,
Eq. (\ref{eq:condpert}) is always satisfied for 
$d=2$, while for $d=1$ one obtains the
restriction $\omega \gg \Delta (\Delta/E_c)$.

For three dimensions Eq. (\ref{tau_eps_pert}) 
can be obtained from the
corresponding dephasing rate 
at $\omega =0$ and $T \tau_{\rm nn} ( T ) \gg 1$ by
simply replacing $T \rightarrow \omega$. For $d \leq 2$ such a procedure
fails. The reason becomes evident
if one goes back to the expressions (\ref{tau_fl_pert})
for the dephasing rate at finite $T$ and zero frequency: 
these equations where derived in the limit $\Gamma_0\gg
1/\tau_{\rm nn}$ but, as already mentioned, the correct $\tau_{\rm nn}$ can still be
obtained for $\Gamma_0 \rightarrow 0$
if $\Gamma_0$ is self-consistently replaced by
$1/\tau_{\rm nn}$. For $T=0$ and $\omega \gg
1 / \tau_{\rm nn}$ it is therefore
reasonable to expect that
both energy-scales $T$ and $\Gamma_0$ must be replaced by
$\omega$ (up to factors of order unity). This 
yields the frequency-dependence given in Eq. (\ref{tau_eps_pert}).

In the following section we shall confirm the above results
by means of a non-perturbative calculation, using the Eikonal expansion of
Ref. \cite{Voelker00}.
\section{Eikonal expansion}
\label{sec:eikonal}
The weak localization correction to the frequency-dependent
conductivity in the presence of Nyquist noise 
can be written as follows  \cite{Voelker00,Voelker00b},
 \begin{eqnarray}
 \delta \sigma ( \omega )
  & = & - \frac{ \sigma_d }{\pi \nu_d } 
 \int_{\tau_{\rm el} }^{\infty} d t e^{  i \omega t } 
 \lim_{T_0 \rightarrow \infty}
 \frac{1}{2 T_0}
 \int_{ - T_0 }^{T_0} dt_0
 \nonumber
 \\
 & \times &
 \langle
 {\cal{C}} ( {\bf{r}} , {\bf{r}} , 
 \frac{t}{2} , - \frac{t}{2} , t_0 ) \rangle
 \; ,
 \label{eq:sigmane}
 \end{eqnarray}
where the Cooperon in the presence of a classical fluctuating scalar potential
$ V ( {\bf{r}} , t )$ satisfies
 \begin{eqnarray}
 \left[  \partial_t  + D {\hat{\bf{P}}}_{\bf{r}}^2  + \Gamma_0
 + i [
 V ( {\bf{r}} , t_0 + t  )  -
 V ( {\bf{r}} , t_0 - t  )  ] \right]
 & & \nonumber
 \\
 & & \hspace{-60mm} \times
 {\cal{C}} ( {\bf{r}} , {\bf{r}}^{\prime} , t , t^{\prime} , t_0 )
 = \delta ( {\bf{r}} - {\bf{r}}^{\prime} )
 \delta ( t - t^{\prime} )
 \; .
 \label{eq:Cooperondef}
 \end{eqnarray}
Here $\hat{\bf{P}}_{\bf{r}} = - i \nabla_{\bf{r}}$ is the
momentum operator, 
and the correlator of the fluctuating scalar potential is
given by the fluctuation-dissipation theorem \cite{Altshuler82,Stern90}, 
 \begin{equation}
 \langle V ( {\bf{q}} , \omega ) V ( {\bf{q}}^{\prime} ,
 \omega^{\prime} ) \rangle =
  2 \pi L^d  
 \delta_{ {\bf{q}} + {\bf{q}}^{\prime} , 0}
 \delta ( \omega + \omega^{\prime} )  
 g ( {\bf{q}} , \omega )
 \; ,
 \label{eq:corphi}
 \end{equation}
where
 \begin{equation}
 g ( {\bf{q}} , \omega ) = - f_{\bf{q}} \coth \left( 
 \frac{\omega}{2T} \right)
 {\rm Im} \epsilon^{-1} ( {\bf{q}} , \omega +i0) 
 \label{eq:gdef}
 \; .
 \end{equation}
Here $\langle \cdots \rangle$ denotes averaging over the 
Gaussian probability distribution of the
Nyquist noise. In the diffusive regime the disorder
averaged dielectric function  $\epsilon ( {\bf{q}} , \omega +i0)$ 
is given by Eq. (\ref{eq:epsdef}).
The solution of Eq. (\ref{eq:Cooperondef}) can be represented as a
Feynman path integral, which was the starting point of the calculations
of Ref. \cite{Altshuler82}. Recently \cite{Voelker00} we have proposed an alternative
way of obtaining the solution of Eq. (\ref{eq:Cooperondef}), which is based on an
Eikonal expansion.  
Following this procedure \cite{Voelker00},
we obtain for  the weak localization correction to the conductivity at finite frequencies,
 \begin{equation}
 \delta \sigma  ( \omega )
 \approx  - \frac{\sigma_d }{\pi \nu_d L^d } 
 \sum_{\bf{k}}
\int_{\tau_{\rm el}}^{\infty} dt 
e^{  - (  D {\bf{k}}^2  - i \omega + \Gamma_0 ) t
 - \Gamma (  t ) }
 \; ,
 \label{eq:sigmawl2}
 \end{equation}
where the Debye-Waller factor is 
\begin{eqnarray}
\label{Gamma_Cooperon}
\Gamma(t) & = & \frac{4}{\nu_d L^d } \sum_{ {\bf{q}}}
\int_{0}^{  \omega_c } \frac{d \omega^{\prime}}{2\pi} \coth \left(
\frac{\omega^{\prime}}{2T} \right)
\frac{\omega^{\prime}}{D{\bf q}^2}
 \nonumber
 \\
 & \times & 
 \left[\frac{D{\bf q}^2}{(D{\bf q}^2)^2+ \omega^{\prime 2}} \left(t - \frac{\sin(\omega^{\prime}
      t)}{\omega^{\prime}} \right) +   
 \right.
  \nonumber \\ 
 & + & \left. 
\text{Re} \frac{e^{-(D{\bf q}^2- i\omega^{\prime})t}-1}{(D{\bf q}^2- i\omega^{\prime})^2} 
- \frac{e^{-D{\bf q}^2t} - \cos(\omega^{\prime} t)}{(D{\bf q}^2)^2+
  \omega^{\prime 2}} \right]
\; .
\end{eqnarray}
The cutoff $\omega_c $ in the frequency-integral is due to the fact that
in Eq. (\ref{eq:Cooperondef}) the scalar potential is treated as 
a classical field, which is only accurate for sufficiently small
frequencies.
The correct choice of
$\omega_c$ has caused some controversy in the recent literature
\cite{Zaikin98,Altshuler98,Aleiner98}. The ``conventional'' way
\cite{Altshuler98,Aleiner98} 
is to choose $\omega_c=\max(T, \omega)$ to explicitly take the
Pauli-principle into account. In contrast,  the authors of
Ref.\cite{Zaikin98} have claimed  that the correct cutoff should be
$\omega_c=1/\tau_{\rm el}$,
which is the frequency that limits the diffusive
regime. If this would be correct, then $1/\tau_{\rm nn}$ would obviously be
independent for $\omega$ for 
$\omega\ll 1/\tau_{ \rm el}$. 
In this work we adopt the conventional choice, so that
$\omega_c =  \omega $
for  $T\ll \omega$. 
Then it is justified to set
 $T=0$  and consequently $\coth(\omega^{\prime}/(2T))=1$. 

\subsection{Bulk systems}

\begin{figure}
\epsfysize4.0cm 
\hspace{5mm}
\epsfbox{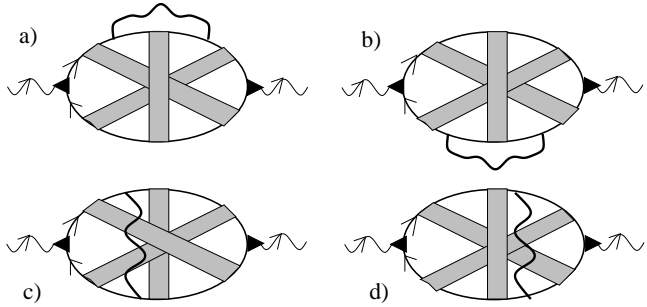}
\vspace{5mm}
\caption{Diagrams contributing to the 
weak localization correction
to the conductivity to first order in the
  screened interaction. The shaded boxes represent the non-interacting
  Cooperon ${\cal C}_0$, the black triangles represent current-vertices,
and  the thick wiggled lines represent the screened interaction.
 The upper (lower) 
solid lines denote the  retarded (advanced) single-particle Green functions. 
Diagrams (a) and (b) are self-energy corrections, while (c)
  and (d) are vertex corrections.
\label{Strom_fig}
}
\end{figure}

Taking the limit $ L \rightarrow \infty$ in Eq. (\ref{Gamma_Cooperon})
and introducing 
the dimensionless variables $x =D{\bf q}^2 t$ and $y = \omega^{\prime} t$, we obtain
\begin{eqnarray}
\label{Gamma_xy}
\Gamma(t)  & = & \frac{ K_d t^{1-d/2}}{\pi\nu_d D^{d/2}}
\int_0^\infty dx \int_0^{\omega t} dy x^{d/2-1} y 
 \nonumber
 \\
 &  \times &
\left\{
  \frac{1}{x^2+y^2} \left[1-\frac{\sin y}{y} \right] 
 \right.
 \nonumber
 \\
 & & 
 + \frac{1}{x(x^2+y^2)} \left[ \left(e^{-x}
    \frac{x^2-y^2}{x^2+y^2}+1 \right) \cos y
\right.   \nonumber 
\\
 &  &
\left.
 \left.  - \frac{x^2-y^2}{x^2+y^2}
  -e^{-x} -\frac{2 x y e^{-x} \sin y}{x^2+y^2} \right] \right\}
 \; .
 \end{eqnarray}
Here
$K_d=2\pi^{d/2}/[(2\pi)^d {\it\Gamma}(d/2)]$ (where ${\it\Gamma}(x)$ is the Gamma
function) is the surface of the
$d$-dimensional unit sphere divided by 
$(2\pi)^{d}$, i.e. $K_1=1/\pi$, $K_2=1/(2\pi)$ and
$K_3=1/(2\pi^2)$. Since $\tau_{\rm nn}$ is
determined from the condition $\Gamma(\tau_{\rm nn}) = 1$,
only the leading asymptotic behavior for
$\omega t \gg 1$ is of interest. The integrals are evaluated by
approximating
$e^{-x}\approx 1-x$ for $x<1$ and $e^{-x}\approx 0$ for $x>1$. The
dominant contribution for $\omega  t \gg 1$ is due to the first
term in the curly braces of Eq. (\ref{Gamma_xy}).
Diagrammatically, this term
corresponds to the self-energy diagrams shown in Fig.\ref{Strom_fig} (a) and (b), so that
for the calculation of the dephasing rate at frequencies $ \omega \gg T$
the vertex corrections in 
Fig.\ref{Strom_fig} (c) and (d) can be ignored.
Note that in dimensions $ d \leq 2$ the vertex corrections are crucial
to obtain the correct result for $1 / \tau_{\rm nn} ( \omega , T  )$ 
in the opposite limit $ \omega \ll T$, see
Eqs. (\ref{eq:tauphi1}) and (\ref{eq:tauphi2}).
Calculating the integrals we finally arrive at 
\begin{equation}
\Gamma(t) \approx \frac{C_d \omega^{d/2}}{\nu_d
  D^{d/2}} t
 \; ,
\end{equation}
with $C_1=\sqrt{2}/\pi , C_2=1/(4\pi)$ and $C_3=1/(6\pi^2)$. The
dephasing rate for $T=0$ and $\omega \tau_{\rm nn}\gg 1$ is thus
given by
\begin{equation}
\label{taunn_phi_eps}
\frac{1}{\tau_{\rm nn} ( \omega ) }=\frac{C_d}{\nu_d}
\left(\frac{\omega}{D}\right)^{{d}/{2}} \; ,
\end{equation}
in agreement with the direct diagrammatic calculation 
presented in  Sec.\ref{sec:diagram}.
Expressing  Eq. (\ref{taunn_phi_eps}) in terms of the
dimensionless conductance $g = E_c / \Delta = D \nu_d L^{d-2} = \sigma_d L^{d-2} / e^2$, 
we may also write
\begin{equation}
\label{taunn_phi_eps2}
\frac{1}{\tau_{\rm nn} ( \omega ) }= C_d \frac{ E_c }{g}
\left(\frac{\omega}{E_c}\right)^{{d}/{2}} \; .
\end{equation}

\subsection{Mesoscopic systems}

Mesoscopic systems are characterized by 
 \begin{equation}
 \frac{1}{\tau_{\phi} (  \omega , T ) } 
 \raisebox{-0.5ex}{$\; \stackrel{<}{\sim} \;$}  \frac{D}{L^2} 
 \; .
 \end{equation}
In this case it is not allowed to replace the  ${\bf{q}}$-summations in 
Eq. (\ref{Gamma_Cooperon}) by integrations.
In the following we assume that the total dephasing rate is dominated by the
Nyquist noise contribution $ 1/ \tau_{\rm nn}$, and consider only
quasi $d$-dimensional samples with
periodic boundary conditions\cite{footnoteboundary}.
Assuming overall charge neutrality, the ${\bf{q}} = 0$-mode
in Eq. (\ref{Gamma_Cooperon}) should be omitted, so that the 
terms $D {\bf{q}}^2$ cannot be smaller than the Thouless energy
$E_c = D / L^2$.
In the regime of interest ($ t 
 \raisebox{-0.5ex}{$\; \stackrel{>}{\sim} \;$} \tau_{\rm nn}$)
all terms in Eq. (\ref{Gamma_Cooperon}) involving the factor
$e^{ - D {\bf{q}}^2 t}$
are exponentially small and can be ignored. 
Choosing again $ \omega_c = \omega$ we obtain for the 
Debye-Waller factor at $T = 0$, 
\begin{eqnarray}
\Gamma(t) & = & \frac{4}{\nu_d L^d } \sum_{ {\bf{q}} \neq 0 }
\int_{0}^{  \omega } \frac{d \omega^{\prime}}{2\pi} \coth \left(
\frac{\omega^{\prime}}{2T} \right)
\frac{\omega^{\prime}}{D{\bf q}^2}
 \nonumber
 \\
 & \times & 
 \left[\frac{D{\bf q}^2}{(D{\bf q}^2)^2+ \omega^{\prime 2}} \left(t - \frac{\sin(\omega^{\prime}
      t)}{\omega^{\prime}} \right)  
 \right.
  \nonumber \\ 
 & - & \left. 
 \frac{   ( D {\bf{q}}^2 )^2  - \omega^{\prime 2} }{(D{\bf q}^2)^2 +  \omega^{\prime 2}} 
+ \frac{\cos(\omega^{\prime} t)}{(D{\bf q}^2)^2+
  \omega^{\prime 2}} \right]
\; .
 \label{eq:gammameso}
\end{eqnarray}
The dominant contribution is again due to the term linear in $t$. 
Defining the frequency-dependent dephasing rate due to Nyquist noise via
$\Gamma ( \tau_{\rm nn} ( \omega ) ) = 1$, we  obtain
 \begin{equation}
 \frac{1}{\tau_{\rm nn} ( \omega ) } = 
\frac{4}{\nu_d L^d} \sum_{ {\bf{q}} \neq 0}
 \int_{0}^{  \omega } \frac{d \omega^{\prime}}{2\pi} \coth \left(
\frac{\omega^{\prime}}{2T} \right)
\frac{\omega^{\prime}}{( D{\bf q}^2)^2 + \omega^{\prime 2} }
\; .
\end{equation}
For $T =0$ and $ \omega \tau_{\rm nn} ( \omega ) \gg 1$ this yields
\begin{equation}
 \frac{1}{\tau_{\rm nn} ( \omega ) } = 
\frac{\Delta}{\pi} \sum_{ {\bf{q}} \neq 0}
\ln \left[ 1 +
\left( 
\frac{\omega}{ D{\bf q}^2 } \right)^2 \right]
\; .
\label{eq:dephres}
\end{equation}
Recall that $\Delta = ( \nu_d L^d )^{-1}$ is the 
level spacing at the Fermi energy.
Eq. (\ref{eq:dephres}) is formally identical with the result for the inverse quasiparticle
dephasing rate derived by Sivan, Imry, and Aronov \cite{Sivan94}
from a phenomenological phase-uncertainty concept. However, in their work
$\omega$ is introduced as the quasiparticle energy, 
whereas $\omega $ in Eq. (\ref{eq:dephres})
is the frequency of the external driving field, which can be controlled experimentally.
For $\omega \gg E_c$ the sum in Eq. (\ref{eq:dephres}) can be converted to an integral
and we recover the bulk result (\ref{taunn_phi_eps}).
In this case $1 / \tau_{\rm nn} ( \omega ) \gg \Delta$, so that the spectrum is
continuous, see Eq. (\ref{condspec}).  
Obviously, this is the case for all frequencies  $ \omega \gg E_c$.
On the other hand, for $ \omega  \ll E_c$ the system is effectively zero-dimensional
with a discrete spectrum. In this case Eq. (\ref{eq:dephres}) reduces to
\begin{equation}
 \frac{1}{\tau_{\rm nn} ( \omega ) } \propto
{\Delta}
\left( 
\frac{\omega}{ E_c } \right)^2 
 \; \; , \; \; 
\omega 
\raisebox{-0.5ex}{$\; \stackrel{<}{\sim} \;$}  E_c
\; .
\label{eq:dephres2}
\end{equation}
As emphasized by Blanter \cite{Blanter96}, in a system with a discrete spectrum
an expansion in terms of the usual Cooperons is uncontrolled, 
and one should use random matrix theory to study the dephasing problem.
Hence, a priori
Eq. (\ref{eq:dephres2})  cannot be trusted. 
Note, however, that  in Ref.\cite{Blanter96}
it was shown that the qualitative behavior of the electron-electron scattering rate
for a system with a discrete spectrum can be obtained
by extrapolating the result derived for a continuous spectrum
to the regime of small frequencies where $ \Delta \tau_{\rm nn} ( \omega ) \ll 1$.
 
\section{Summary and conclusions}

In this work we have calculated
the frequency-dependence of the
dephasing-rate  due to Nyquist noise, $1/\tau_{\rm nn}(\omega, T )$,
in the regime $\omega \gg \mbox{max} \{ T, 1 / \tau_{\rm nn} \}$.
Our main result is given in Eq. (\ref{taunn_phi_eps}),
which shows that $ 1 / \tau_{\rm nn} ( \omega , 0 )$
exhibits for all dimensions 
$d$ the same frequency-dependence
as the inelastic quasiparticle damping rate
$ 1 / \tau_{\rm ee} ( \omega )$
due to electron-electron scattering \cite{Altshuler85}. 
By examining the perturbative expression for
$1 / \tau_{\rm nn} ( \omega )$ given in Eq. (\ref{tau_rpa2}),
it is easy to see that the dominant contribution  
is due to frequencies $\omega^{\prime}$ 
of the order of the external frequency $\omega$. 
Similarly, 
 the inelastic quasiparticle scattering rate $1/\tau_{\rm ee}( \omega )$ 
is governed by frequency transfers of order $\omega$ \cite{Altshuler85}.
In contrast,  the dephasing rate $1 / \tau_{\rm nn} ( \omega , T )$ 
for $\omega \ll T$ is dominated by frequency 
transfers of order $1/\tau_{\rm nn}$. 
As a result, in $d \leq 2$ the frequency-dependence of 
$1 / \tau_{\rm nn} ( \omega , 0)$ cannot be obtained from 
the temperature-dependence 
of $1 / \tau_{\rm nn} ( 0 , T )$
by  simply replacing $ T \rightarrow \omega$.

Finally, let us discuss the relevance of our result for experiments.
The authors of Ref.\cite{Altshuler98} have 
proposed that the experimentally observed saturation
of the dephasing rate is due to external microwave radiation which is not completely
shielded from the sample. For sufficiently small frequencies, $ \omega \ll
1 / \tau_{\rm AC} ( \omega )$, the corresponding non-equilibrium dephasing rate
$1 / \tau_{\rm AC} ( \omega )$ is proportional to $\omega^{2/5}$ \cite{Altshuler81}, see
Eq. (\ref{eq:tauac}).
At higher frequencies $1 / \tau_{\rm AC} ( \omega )$ vanishes as
$\omega^{-2}$.
The crossover frequency $\omega_{\ast}$ separating
the low-frequency from the high-frequency regime has been estimated to
be in the GHz-range for experimentally relevant parameters \cite{Altshuler98}.
Obviously, for $ \omega \gg \omega_{\ast}$
the contribution from the 
non-equilibrium dephasing rate $1 / \tau_{\rm AC} ( \omega )$ 
to the total dephasing rate
in Eq. (\ref{eq:tautotal})  is negligible, so that the 
dephasing rate at $T = 0$ should be 
dominated by the frequency-dependence of the
Nyquist-noise contribution,
 \begin{equation}
 \frac{1}{\tau_{\phi}^{\rm tot} ( \omega)} \approx \frac{1}{\tau_{\rm nn} ( \omega )}
 \approx C_d  \frac{E_c}{g} \left( \frac{\omega}{E_c} \right)^{d/2}
 \; .
 \end{equation} 
In particular, in two dimensions at $T = 0$
 \begin{equation}
 \frac{1}{\tau_{\phi}^{\rm tot} ( \omega)} \approx 
 \frac{\omega}{4 \pi g}
 \; \; , \; \;  d = 2 
 \; ,
 \end{equation}
and in one dimension
 \begin{equation}
 \frac{1}{\tau_{\phi}^{\rm tot} ( \omega)} \approx 
 \frac{ \sqrt{2 \omega E_c} }{ \pi g}
 \; \; , \; \;  d = 1 
 \; .
 \end{equation}
Note that for  good two-dimensional metals  (with 
typical values of $g$ in the regime
$10 \raisebox{-0.5ex}{$\; \stackrel{<}{\sim} \;$} g
\raisebox{-0.5ex}{$\; \stackrel{<}{\sim} \;$} 100$) the condition
$\omega \gg 1 / \tau_{\rm nn} ( \omega )$ is always satisfied.
For frequencies  $\omega / ( 2 \pi)$ of a few  GHz one should then observe
dephasing times  in the nano-second range. 
Moreover, in $d=2$ the product $ \omega \tau_{\rm nn} ( \omega )$
should only depend on  the
dimensionless conductance $g$. 
We expect that
our predictions can be  checked  
with the available experimental techniques \cite{Mohanty97,Gougam00}.
In practice, it is not so easy to feed GHz microwaves
into a mesoscopic metallic sample in a well-defined manner, because
the fields can be non-uniform due to the skin effect and the
electrodynamics near the contacts is difficult to
predict. 
However, it seems that  these problems
are not unsurmountable. For example,
the non-uniformity of the field inside the sample can be minimized
by using sufficiently thin samples  \cite{Liu91,Vitkalov96},  
with thickness smaller than the skin depth.
Furthermore,
with the help of 
contactless measuring techniques
uncontrollable stay capacitances due to 
leads can be avoided  \cite{Vitkalov96}.

We conclude that AC measurements of the weak localization correction
to the conductivity can shed  new light on the mechanism for the
observed saturation of the dephasing rate in low-dimensional metals.

This work was supported by the DFG 
via Project No. KO 1442/3-1 and SFB 345 at 
G\"{o}ttingen University.

\begin{thebibliography}{99}
\bibitem[*]{address}
Present address:
IBM Deutschland GmbH, Laatzener Strasse 1, 
30539 Hannover, Germany.
%
\bibitem{Altshuler85}
B.L. Altshuler and A.G. Aronov in {\it Electron-Electron Interactions
  in Disordered Systems}, edited by A.L. Efros and M. Pollak (North
Holland, Amsterdam, 1985). 
%
\bibitem{Altshuler82}
B. L. Altshuler, A. G. Aronov, and D. E. Khmelnitskii,
J. Phys. C {\bf{15}}, 7367 (1982).
%
\bibitem{Voelker00}
A. V\"olker and P. Kopietz, Phys. Rev. B {\bf 61}, 13508 (2000).
%
\bibitem{Mohanty97}
P. Mohanty, E.M.Q. Jariwala, and R.A. Webb, Phys. Rev. Lett. {\bf 78},
3366 (1997).
%
\bibitem{Altshuler98}
B. L. Altshuler, M. E. Gershenson, and I. Aleiner,
Physica E {\bf{3}}, 58 (1998).
%
\bibitem{Zaikin98}
D. S. Golubev and A. D. Zaikin, Phys. Rev. Lett. {\bf{81}}, 1074 (1998);
D. S. Golubev and A. D. Zaikin, Phys. Rev. B {\bf 59}, 9195 (1999).
%
\bibitem{Imry99} 
Y. Imry, H. Fukuyama, and P. Schwab, Europhys. Lett. {\bf{47}}, 608 (1999).
%
\bibitem{Zawadowski99}
A. Zawadowski, Jan von Delft, and D. C. Ralph, Phys. Rev. Lett.
{\bf{83}}, 2632 (1999).
%
\bibitem{Altshuler81}
B. L. Altshuler, A. G. Aronov, and D. E. Khmelnitskii,
Solid State Commun. {\bf{39}}, 619 (1981).
%
\bibitem{Aleiner98}
I. L. Aleiner, B. L. Altshuler, and M. E. Gershenson, Phys. Rev.Lett. {\bf{82}}, 3190 (1999);
cond-mat/9808053.
%
\bibitem{Voelker00b}
A. V\"{o}lker, PhD-Thesis (Universit\"{a}t G\"{o}ttingen, 2000, http://webdoc.sub.gwdg.de/diss/2000/voelker/index.htm).
%
\bibitem{Stern90}
A. Stern, Y. Aharonov, and Y. Imry, Phys. Rev. A {\bf{41}}, 3436 (1990);
Sh. Kogan, {\it{Electronic Noise and Fluctuations in Solids}},
(Cambridge University Press, Cambridge, 1996).
%
\bibitem{footnoteboundary}
Qualitatively, the behavior of the dephasing rate is not
sensitive to the boundary conditions, see Ref.\cite{Blanter96}. 
%
\bibitem{Blanter96}
Y. M. Blanter, Phys. Rev. B {\bf{54}}, 12807 (1996).
%
\bibitem{Sivan94}
U. Sivan, Y. Imry, and A. G. Aronov, Europhys. Lett. {\bf{28}}, 115 (1994).
%
\bibitem{Gougam00}
A. B. Gougam, P. Pierre, H. Pothier, D. Esteve, and N. O. Birge,
J. Low Temp. Phys. {\bf{118}}, 447 (2000);
D. Natelson, R. L. Willett, K. W. West, and L. N. Pfeiffer,
Phys. Rev.Lett. {\bf{86}}, 1821 (2001).
%
\bibitem{Liu91}
J. Liu and N. Giordano, Phys. Rev. B {\bf{43}}, 1385 (1991).
%
\bibitem{Vitkalov96}
S. A. Vitkalov, Zh. Eksp. Teor. Fiz. {\bf{109}}, 1846 (1996) [Sov. Phys. JETP
{\bf{82}}, 994 (1996)].

\end{thebibliography}
\end{document}